\documentstyle[11pt]{article}
\newcommand{\lang}[1]{}     % nur fuer die Lang-Version
\newcommand{\unit}[1]{\mbox{\ #1}}

\newcommand{\comment}[1]{}

\newcommand{\FA}{F_{A}}
\newcommand{\FV}{F_{V}}
\newcommand{\mpi}{m_{\pi}}
\newcommand{\mt}{m_{\tau}}
\newcommand{\fpi}{f_{\pi}}
\newcommand{\mcal}{{\cal M}}
\newcommand{\wz}{\sqrt{2}}
\newcommand{\be}{\begin{equation}}
\newcommand{\ee}{\end{equation}}
\newcommand{\gm}{\gamma_{-}}
\newcommand{\gp}{\gamma_{+}}

\newcommand{\gnr}{\Gamma_{\tau \rightarrow \nu \pi}}
\newcommand{\fpg}{F_{\pi \gamma}}
\newcommand{\faktor}{G_{F} \cos \theta_{c} e}
\newcommand{\Res}{\mbox{Res}}
\newcommand{\rp}{{\rho'}}
\newcommand{\rpp}{{\rho''}}
\newcommand{\mr}{m_{\rho}}
\newcommand{\mrp}{m_{\rp}}
\newcommand{\mrpp}{m_{\rpp}}
\newcommand{\ma}{m_{a_{1}}}
\newcommand{\ga}{\Gamma_{a_{1}}}

\newcommand{\br}{\mbox{BR}}
\newcommand{\strich}[1]{#1  \! \! \slash}
\newcommand{\gapg}{\Gamma_{a_{1} \rightarrow \pi \gamma}}
\newcommand{\grpg}{\Gamma_{\rho \rightarrow \pi \gamma}}
\newcommand{\grppo}{\Gamma_{\rho' \rightarrow \pi \omega}}

\renewcommand{\Re}{\mbox{Re}}

\begin{document}
%--------------------------------------
%-----------------------------
\everymath={\displaystyle}
\thispagestyle{empty}
\vspace*{-2mm}
%\begin{center}
\thispagestyle{empty}
\noindent
\hfill TTP93--1\\
\mbox{}
\hfill  February 1993  \\
%                 \hfill \today \\
\vspace{0.5cm}
\begin{center}
  \begin{Large}
  \begin{bf}
RADIATIVE TAU DECAYS\\
WITH ONE PSEUDOSCALAR MESON
   \\
  \end{bf}
  \end{Large}
  \vspace{0.8cm}
  \begin{large}
   Roger Decker and Markus Finkemeier\\[5mm]
    Institut f\"ur Theoretische Teilchenphysik\\
    Universit\"at Karlsruhe\\
    Kaiserstr. 12,    Postfach 6980\\[2mm]
    7500 Karlsruhe 1, Germany\\
  \end{large}
  \vspace{4.5cm}
  {\bf Abstract}
\end{center}
\begin{quotation}
\noindent
We have calculated the decay $\tau \rightarrow \nu \pi(K) \gamma$.
We present the photon energy spectrum,
the meson-photon invariant mass
spectrum and
the integrated rate
as a function of a photon energy cut or an invariant mass cut.
Both the
internal bremsstrahlung and the structure dependent radiation
have been taken into account. To this aim we have
parametrized the form factors $F_V$ and
$F_A$, which determine the structure dependent radiation.
Observables especially suited for the measurement of the
structure dependent form factors are found and implications on
the width of the $a_1$ discussed.
\end{quotation}
%-----------------------------------------------------------------------
\newpage
\setcounter{page}{2}
%====================================================

%========================================================
\section{Introduction}
Present and future high luminosity electron positron colliders  and
especially tau-charm factories will produce very large samples of
tau pairs in the near future. This will give the opportunity to
observe rare tau decays, for example radiative semihadronic modes, the
simplest of which are $\tau \rightarrow \nu \pi \gamma$ and
$\tau\to\nu K \gamma$,
which
we will therefore calculate in this paper.

The decay $\tau\to\nu\pi\gamma$
is related by crossing symmetry to the radiative pion
decay ($\pi \rightarrow e \nu \gamma$), where the electron plays the
role of the tau lepton. There is, however, a crucial difference,
namely that the momentum transfer squared $t$
between the leptonic and the pion-photon system,
which is almost zero in the case of the pion decay, can take any
value up to $\mt^2$ in the radiative tau decay. Several resonances,
which can contribute only as small virtual corrections in the pion decay,
can now be produced as real particles and duly have to be taken
into account.

The decay amplitude can be divided into internal bremsstrahlung,
 where the pion behaves like a pointlike particle and which
therefore is completely
determined by quantum electrodynamics, and the structure dependent
part, which is governed by strong interaction effects and is
parametrized by two formfactors $F_V(t)$ and $F_A(t)$. So the main task
is to obtain these formfactors. We will assume them to be dominated
by several Breit Wigner type resonances and determine their parameters
 by employing different constraints, both theoretical and experimental
ones. For instance normalizations at $t=0$ are fixed by the
Adler-Bell anomaly
($F_V$) and the $\pi\to e\nu\gamma$ measurement ($F_A$), respectively.
On the other hand the relative strengths of the different resonances
are taken from a high energy QCD prediction and experimental data
on the formfactors at finite values of $t$.

This paper is organized as follows: In Sec.\ 2 we shortly discuss
the kinematics, the differential decay rate and lay down our conventions.
In Sec.\ 3 the general form of the matrix
element, photon spectrum and the decay rate of $\tau\to\nu\pi\gamma$
are calculated. Their
dependence on the two formfactors $F_V(t)$ and $F_A(t)$ are worked out.
In Sec.\ 4 these formfactors are parametrized
along the lines indicated above. In Sec.\ 5 the numerical results for
the photon and invariant mass spectra and the decay rate are presented.
In Sec.\ 6 we estimate the decay $\tau\to\nu K \gamma$ and in Sec.\ 7
we discuss our results.

%=======================================================================
\section{Kinematics and the Decay Rate}
For definiteness and without loss of generality we consider
the decay of the
negatively charged tau into a pion, a neutrino and a photon
\be
   \tau^-(s) \longrightarrow \nu_\tau(q) \pi^-(p) \gamma(k)
\ee
only
and transcribe the formulae to the case of the
decay into a kaon in Sec. 7.
The kinematics of this decay is equivalent to that of the radiative
pion decay \cite{Bro64}. We can write the differential decay rate
for an unpolarized tau
in such a way that it depends
on two Lorentz scalars, taken to be $x$ and $y$:
\be
    x := \frac{2 s \cdot k}{\mt^2} \mbox{\hspace{2cm}}
    y := \frac{2 s \cdot p}{\mt^2}
\ee
In the tau rest frame $x$ and $y$ are the energies $E_\gamma$ and
$E_\pi$
of the photon and the pion, respectively, expressed in units
of $\mt/2$:
\be
   E_\gamma = \frac{\mt}{2} x \mbox{\hspace{2cm}}
   E_\pi    = \frac{\mt}{2} y
\ee
The kinematical boundaries for
 $x$ and $y$ are given by
\begin{eqnarray}
   0 & \leq x  \leq & 1 - r^2
   \nopagebreak \nonumber \nopagebreak \\ \nopagebreak
   1 - x + \frac{r^2}{1 -x} & \leq y \leq & 1 + r^2
\end{eqnarray}
where
\be
   r^2 := \left( \frac{\mpi}{\mt} \right)^2 =
6.17 \times 10^{-3} \ll 1
\ee
using the central value of the new $\tau$ mass
experiments  \cite{taumasse}
\be
   \mt = 1777 \unit{MeV}.
\ee
Let us quote some useful formulae
which relate different kinematical expressions:
\be
   p \cdot k = \frac{\mt^2}{2} (x + y - 1 - r^2)
\ee \be
   t := (s - q)^2 = (k + p)^2 = \mt^2 (x + y - 1)
\ee
Now we are in position to present the formula for the decay rate
of an unpolarized tau decaying into $|\nu_\tau\pi\gamma\rangle$
The differential decay rate is given by
\be
  d\Gamma(\tau \rightarrow \nu_\tau \pi \gamma) =
  \frac{1}{512 \pi^5 E_{\tau}} \delta^{(4)} (k + p + q - s)
  \overline{| \mcal |^2} \frac{d^3 \vec{k} d^3 \vec{p} d^3 \vec{q}}%
  {E_{\gamma} E_{\pi} E_{\nu}}
\ee
where the bar over the matrix element denotes summing over
the photon polarization and neutrino
spin and averaging over the tau spin.

Choice of the tau rest frame, integration over the
neutrino momentum and the remaining angles and introduction of
the dimensionless variables $x$ and $y$ lead to
\be
   \frac{d^2 \Gamma}{dx\, dy} = \frac{\mt}{256 \pi^3}
   \overline{| \mcal |^2}
\ee
The integration over $y$ yields the photon spectrum
\be
    \frac{d\Gamma}{dx} = \int_{1 - x +
\frac{r^2}{1-x}}^{1 + r^2} dy \,
    \frac{d^2 \Gamma}{dx\, dy}
\ee
Because of the infrared divergence of the internal bremsstrahlung
a low energy cut must be introduced for the photon energy, eg. by
requiring
\be
       x \geq x_0
\ee
The integrated decay rate is then obtained by
\be
   \Gamma (x_0) =
   \Gamma (E_0) =
   \int_{x_{0}}^{1 - r^{2}} dx
            \frac{d \Gamma}{dx}
\ee
and does depend on the cut in the photon energy
($E_0 = \frac{\mt}{2} x_0$).

Instead of the invariants $x$ and $y$ we will also use
$x$ and $z$ where $z$ is the scaled momentum transfer
squared:
\be
   z = \frac{t}{\mt^2} = x + y - 1
\ee
Their kinematical boundaries are given by
\begin{eqnarray}
   z - r^2 & \leq x  \leq & 1 - \frac{r^2}{z}
   \nonumber \\ \nopagebreak[4]
   r^2 & \leq z \leq & 1
\end{eqnarray}
Integration of $d^2 \Gamma/dx\, dy$ over $x$ then yields the
spectrum in $z$, ie. the spectrum in the invariant mass
of the pion-photon system:
\be
   \frac{d\Gamma}{dz} (z) =
   \frac{d\Gamma}{dz} (\sqrt{t}) =
   \int_{z - r^2}^{1 - r^2/z} dx
   \frac{d^2 \Gamma}{dx\, dy} (x, y=z-x+1)
\ee
The integrated rate for events with $t \geq t_0$ is then
given by
\be
   \Gamma(z_0) = \Gamma(\sqrt{t_0}) =
   \int_{z_0}^{1} dz \frac{d\Gamma}{dz} (z)
\ee
where, of course, $z_0 = \frac{t_0}{\mt^2}$.
Note that the cut in $t$ acts as both an infrared and a
collinear cut off.

%=======================================================================
\section{General Structure of the Matrix Element
and the Decay Rate}
In complete analogy to the case of the
radiative pion decay \cite{Bae67}, the
matrix element for the decay $\tau \longrightarrow \nu \pi \gamma$
can be written as the sum of four contributions:
\be
   \mcal [\tau^-(s) \longrightarrow \nu_\tau(q) \pi^-(p) \gamma(k)]
   = \mcal_{IB_{\tau}} + \mcal_{IB_{\pi}} + \mcal_{V} + \mcal_{A}
\ee
with
\begin{eqnarray}
   \mcal_{IB_{\tau}} & = & - \faktor \fpi p_\mu
\epsilon_\nu(k) L^{\mu \nu}
   \nonumber \\
   \mcal_{IB_{\pi}} & = & \faktor \fpi \epsilon^\nu(k)
\left( \frac{2p_\nu  (k + p)_\mu}
      {\mpi^2 - t} - g_{\mu\nu} \right) L^\mu \nonumber \\
   \mcal_{V} & = &   \frac{\faktor}{\wz} \frac{\FV(t)}{\mpi}
i\epsilon_{\mu \nu \rho \sigma}
     \epsilon^\nu(k)  k^\rho p^\sigma  L^\mu  \nonumber \\
   \mcal_{A} & = & - \frac{\faktor}{\wz}\frac{\FA(t)}{\mpi}
\epsilon^\nu(k) \left(
(p_\nu k_\mu - (p \cdot k) g_{\mu\nu} \right) L^\mu
\end{eqnarray}
where $G_F$ is the Fermi constant, $\theta_C$ the Cabibbo angle, $e$
the electric charge and $\epsilon_\nu$ the polarisation vector of the
photon. For our numerical results we use the pion
coupling constant $f_\pi$ to be
\cite{Hol90,Rev92}
\be
   \fpi = 92.5 \unit{MeV}
\ee
$F_V$ and $F_A$ are the so called structure dependent form factors which
are parametrized in the next section.
Note that  our definition  implies that these form factors are
dimensionless.
Finally $L^\mu$ and $L^{\mu\nu}$
are leptonic currents defined by
\begin{eqnarray}
   L^\mu & = & \bar{u}(q) \gamma^\mu \gm u(s) \nonumber \\
   L^{\mu \nu} & = & \bar{u}(q) \gamma^\mu \gm
\frac{\strich{k} - \strich{s}
      - \mt}{(k - s)^2 - \mt^2} \gamma^\nu u(s)
\end{eqnarray}
In fact the four terms correspond to the Feynman diagrams in Fig.\ 1:
\begin{itemize}
\item $\mcal_{IB_{\tau}}$ is the bremsstrahlung off the tau (a),
\item $\mcal_{IB_{\pi}}$ the sum of the
pion bremsstrahlung (b) and the seagull (c),
\item $\mcal_{V}$ is the structure
dependent vector (e) and
\item $\mcal_{A}$ the structure dependent axial
contribution (f).
\end{itemize}
Note that all ambiguities due to the strong interactions have been
parametrized in terms of the two form factors $\FA$ and $\FV$.
In fact these form
factors are the same functions of
the momentum transfer $t$
as those in the radiative pion decay, the
only difference being that $t$ now varies from
$0$ up to $\mt$ rather than just up to $\mpi$.

The two matrix elements $\mcal_{IB_{\tau}}$ and $\mcal_{IB_{\pi}}$
are not separately gauge invariant, but their sum, ie. the (total)
matrix element for internal bremsstrahlung IB
\be
   \mcal_{IB} = \mcal_{IB_{\tau}} + \mcal_{IB_{\pi}}
\ee
is, as are $\mcal_V$ and $\mcal_A$. We also define the (total)
structure dependent radiation SD, viz. by
\be
   \mcal_{SD} = \mcal_{V} + \mcal_{A}
\ee
Using standard technics the matrix elements may be simplified, the main
rearrangement being
\be
   - \bar{u}(q) \gp \strich{p}
\frac{\strich{k} - \strich{s} - \mt}{(k-s)^2
      - \mt^2} \strich{\epsilon} u(s)
 = \bar{u}(q) \gp \left[ \mt
\frac{\strich{k} \strich{\epsilon}}{2 s \cdot k}
      - \mt
\frac{s \cdot \epsilon}{s \cdot k} + \strich{\epsilon} \right]
      u(s)
\ee
We obtain
\begin{eqnarray}
   \mcal_{IB} & = & \faktor \fpi \mt \bar{u}(q) \gp \left[
      \frac{p \cdot \epsilon}{p \cdot k} + \frac{\strich{k}
      \strich{\epsilon}}{2 s \cdot k} -
\frac{s \cdot \epsilon}{s \cdot k}
      \right] u(s) \nonumber \\
   \mcal_{SD} & = & \frac{\faktor}{\wz} \left\{ i \epsilon_{\mu \nu \rho
      \sigma} L^\mu \epsilon^\nu k^\rho p^\sigma
\frac{\FV(t)}{\mpi}
      + \bar{u}(q) \gp \left[ (p \cdot k) \strich{\epsilon} -
      (\epsilon \cdot p) \strich{k} \right] u(s)
\frac{\FA(t)}{\mpi}
      \right\}
\end{eqnarray}
The square of the matrix element is then given by
\be
  \overline{| \mcal |^2} = \overline{| \mcal_{IB} |^2}
  + 2 \overline{\Re (\mcal_{IB} \mcal_{SD}^\star)}
  + \overline{| \mcal_{SD} |^2},
\ee
where --- as stated in the previous section ---
the bar denotes summing over the photon polarization and neutrino
spin and averaging over the tau spin.

In order to make our results transparent we have divided
the decay rate into several parts:
\begin{itemize}
\item the internal bremsstrahlung part
$\Gamma_{IB}$ arising
from $\overline{| \mcal_{IB} |^2}$,
\item the structure dependent part
$\Gamma_{SD}$ arising from $\overline{| \mcal_{SD} |^2}$
\item  the
interference part $\Gamma_{INT}$ arising from
$2 \overline{\Re (\mcal_{IB} \mcal_{SD}^\star)}$.
\end{itemize}
Furthermore we have subdivided $\Gamma_{SD}$ in the obvious way into
$\Gamma_{VV}$ arising from the vector matrix element squared,
$\Gamma_{AA}$ form the axial part and the vector-axial interference
term $\Gamma_{VA}$ and similarly $\Gamma_{INT}$ gets divided into the
internal bremsstrahlung-vector interference $\Gamma_{IB-V}$ and
the internal bremsstrahlung-axial interference
$\Gamma_{IB-A}$.

So our master formulae are:
\begin{eqnarray}
    \Gamma_{total} & = & \Gamma_{IB} + \Gamma_{SD} +
       \Gamma_{INT} \nonumber \\
    \Gamma_{SD} & = & \Gamma_{VV} + \Gamma_{VA} + \Gamma_{AA}
     \nonumber \\
    \Gamma_{INT} & = & \Gamma_{IB-V} + \Gamma_{IB-A}
\end{eqnarray}
One finds for the differential decay rate
\begin{eqnarray}
   \frac{d^2\Gamma_{IB}}{dx \, dy} & = & \frac{\alpha}{2 \pi}
   f_{IB}(x,y,r^2) \frac{\gnr}{(1-r^2)^2} \nonumber
   \\
   \frac{d^2\Gamma_{SD}}{dx \, dy} & = & \frac{\alpha}{16 \pi}
   \frac{\mt^4}{\fpi^2 \mpi^2}
   \left[ |F_V|^2 f_{VV}(x,y,r^2) + 2 \Re (F_V F_A^\star)
f_{VA}(x,y,r^2)
   + |F_A|^2 f_{AA}(x,y,r^2) \right] \frac{\gnr}{(1-r^2)^2} \nonumber
   \\
   \frac{d^2\Gamma_{INT}}{dx \, dy} & = &
\frac{\alpha}{2 \sqrt{2} \pi}
   \frac{\mt^2}{\fpi \mpi}
   \left[ f_{IB-V}(x,y,r^2) \Re(F_V) +
f_{IB-A}(x,y,r^2) \Re(F_A) \right]
   \frac{\gnr}{(1-r^2)^2}
\end{eqnarray}
where
\begin{eqnarray}
   f_{IB} (x,y,r^2) & = & \frac{[r^4 (x + 2) - 2 r^2 (x + y) +
      (x + y - 1)(2 - 3x + x^2 + xy)](r^2 - y + 1)}
{(r^2 - x - y +1)^2
       x^2}
   \nonumber \\
   f_{VV} (x,y,r^2) & = & - [r^4 (x + y) + 2 r^2 (1 - y) (x + y)
      + (x + y - 1)(-x + x^2 - y + y^2)]
   \nonumber \\
   f_{AA} (x,y,r^2) & = & f_{VV}(x,y,r^2)
   \nonumber \\
   f_{VA}(x,y,r^2) & = & [r^2 (x + y) + (1 - x - y)(y-x)]
      (r^2 - x - y + 1)
   \nonumber \\
   f_{IB-V}(x,y,r^2) & = & - \frac{(r^2 - x - y + 1)(r^2 - y + 1)}{x}
   \nonumber \\
   f_{IB-A}(x,y,r^2) & = & \frac{[r^4 - 2 r^2(x + y) + (1 - x + y)
      (x + y - 1)](r^2 - y + 1)}{(r^2 - x - y + 1) x}
\end{eqnarray}
In the approximation $r^2 \approx 0$ (vanishing pion mass) the formulae
simplify to
\begin{eqnarray}
   \label{eqnib2}
   f_{IB} (x, y, 0) & = &  \frac{[ 1+(1-x)^2 -x(1-y)](1-y)}%
      {(x + y - 1) x^2}
   \nonumber \\
   f_{VV} (x,y,0) & = & (x - x^2 + y - y^2)(x + y - 1)
   \nonumber \\
   f_{VA}(x,y,0) & = & (x + y - 1)^2 (y - x)
   \nonumber \\
   f_{IB-V}(x,y,0) & = & \frac{(x + y - 1)(1 - y)}{x}
   \nonumber \\
   f_{IB-A}(x,y,0) & = & \frac{(x - y - 1)(1 - y)}{x}
\end{eqnarray}
Note that we expressed the radiative decay rate in terms of the rate of the
non-radiative decay ($\tau \to \nu_\tau \pi$)):
\be
   \gnr = \frac{G_F^2 \cos^2 \theta_c \fpi^2}{8 \pi} \mt^3
      (1 - r^2)^2
\ee
Although the factor  $\mt^2 / (\fpi \mpi)$ in front of
$\FA$ and $\FV$  is rather large a number ($ \approx 240$),
we should mind that the form factors $F_A$ and $F_V$
themselves are of the order of
$10^{-2}$ (see next section), so that we expect  the different
contributions
to be of the same order of magnitude.
Note also that the internal bremsstrahlung and the interference term
are infrared divergent in the limit $x \rightarrow 0$
(vanishing photon energy). Furthermore there is an enhancement of
the internal bremsstrahlung (but not of the interference term) in
the region
\be
  (x + y - 1) \rightarrow r^2 \; \Leftrightarrow \;
  t \rightarrow \mpi^{2}
\ee
This enhancement is due to large logs of the pion mass,
the usual collinear
divergencies for  massless particles.

Next we would like to make a short comment on
the internal bremsstrahlung part of the decay rate.
Let us first give the corresponding expression
in the radiative pion decay \cite{Bry82} (neglecting $m_e$)
\be
   \frac{d^2 \Gamma_{IB}(\pi \rightarrow e \nu \gamma)}
   {dx \, dy} = \frac{\alpha}{2 \pi}
   \Gamma(\pi \rightarrow e \nu) IB(x,y)
\ee
with
\be
   \label{eqnib1}
   IB(x,y) = \frac{(1 - y)[1 + (1 - x)^2]}{x^2 (x + y - 1)}
\ee
where now, however, $x$ and $y$ are the photon and electron energies
in units of $\mpi / 2$:
\be
   E_\gamma = \frac{\mpi}{2} x \mbox{\hspace{2cm}}
   E_e    = \frac{\mpi}{2} y
\ee
Expressed  in this form eq.(\ref{eqnib1}) is identical with
eq.(\ref{eqnib2}),
apart from the $x(1-y)$
term. On the other hand keeping in mind the definition of $x$
we note that $\mt / 2$ and $m_\pi / 2$, respectively, set the scale
for photons
to be considered as ``hard'' or ``soft''. This means that
the formulae for internal bremsstrahlung should be similar for
radiative tau and pion decay, once they are expressed in terms of
$x$ and $y$, which is indeed the case.
So photons from the $\tau$- or $\pi$- decays of comparable ``softness'',
e.g. comparable $x$, have very different energies.

Finally we present analytical expressions for the
invariant mass spectrum:
\begin{eqnarray}
   \frac{d\Gamma_{IB}}{dz} & = &
   \frac{\alpha}{2 \pi} \Big[ r^4 (1 -z) + 2 r^2 (z - z^2)
   - 4z + 5 z^2 - z^3 +
\nonumber \\
   &  & + (r^4 z + 2 r^2 z - 2z -2z^2 + z^3) \ln z \Big]
   \frac{1}{z^2 - r^2 z}
   \frac{\gnr}{(1 - r^2)^2}
\nonumber \\
   \frac{d\Gamma_{VV}}{dz} & = &
   \frac{\alpha}{48 \pi} \frac{\mt^4}{\fpi^2 \mpi^2}
   \frac{(z - 1)^2 (z - r^2)^3 (1 + 2z)}{z^2}
   |F_V|^2
   \frac{\gnr}{(1 - r^2)^2}
\nonumber \\
   \frac{d\Gamma_{VA}}{dz} & = & 0
\nonumber \\
   \frac{d\Gamma_{AA}}{dz} & = &
   \frac{\alpha}{48 \pi} \frac{\mt^4}{\fpi^2 \mpi^2}
   \frac{(z - 1)^2 (z - r^2)^3 (1 + 2z)}{z^2}
   |F_A|^2
   \frac{\gnr}{(1 - r^2)^2}
\nonumber \\
   \frac{d\Gamma_{IB-V}}{dz} & = &
   \frac{\alpha}{2 \sqrt{2} \pi}
   \frac{\mt^2}{\fpi \mpi}
   \frac{(z-r^2)^2 (1 -z  + z \ln z)}{z} \Re(F_V)
   \frac{\gnr}{(1 - r^2)^2}
\nonumber \\
   \frac{d\Gamma_{IB-A}}{dz} & = &
   \frac{\alpha}{2 \sqrt{2} \pi}
   \frac{\mt^2}{\fpi \mpi}
   \Big[ r^2 (1 - z) - 1 -z + 2 z^2 +
 \nonumber \\
   & & + (r^2 z - 2z - z^2) \ln z \Big]
   \frac{z-r^2}{z} \Re(F_A)
   \frac{\gnr}{(1 - r^2)^2}
\end{eqnarray}
Note that the interference
terms IB-V and IB-A are now finite in the limit $z \to r^2$, which
proves that their infrared divergencies are integrable.

%=======================================================================
\section{Parametrization of the Form Factors}
\subsection{The Axial Form Factor}
The axial form factor $F_A(t)$ fulfills a once subtracted dispersion
relation. Since this form factor is measured at $t=0$ in the radiative
pion decay we will use this value to fix the subtraction constant.
Furthermore it is supposed to be dominated by the $a_1$ meson.
The following form includes these constraints:
\be
    F_A(t) = F_A(0) \Res_{a_1}(t)
    \label{eqnfa}
\ee
where
$F_A(0)$ is taken from The Particle Data Group \cite{Rev92}
\be
   F_A(0)_{exp} = -0.0116 \pm 0.0016.
   \label{eqn52}
\ee
and  $\Res_{a_1}(t)$ is a normalized Breit Wigner resonance factor
\be
   \Res_{a_1}(t = 0) = 1
\ee
Note that most experiments on the radiative pion decay have
been performed with pions at rest. These experiments are
sensitive mainly to
$(1 + \gamma)^2$ only, where
\be
   \gamma = \frac{F_A(0)}{F_V(0)}
\ee
So they also allow for a positive sign solution for $F_A(0$:
\be
   F_A(0) = + 0.0422
   \label{eqn53}
\ee
Measurements performed with pions in flight support
our value in (\ref {eqn52}). However, we will also comment on implications
of the other
solution.

Since the decay of the $a_1$ is strongly dominated by its three pion decay
we approximate  in the Breit Wigner the total width by the three pion
width. Furthermore we will use a energy dependent width as
given in \cite{Kue90}
\be
     \Res_{a_1}(t) = \frac{\ma^2 }%
        {\ma^2 - t - i \ma \Gamma_{ks}(t)}
\ee
where
\be
   \Gamma_{ks}(t) = \frac{g(t)}{g(\ma^2)} \Gamma_{a_1}
\ee
with
\be
   g(t) = \left\{ \begin{array}{lll}
      0 & \mbox{\ \ if\ \ } t < 9 \mpi^2 \\
      4.1 (t - 9 \mpi^2)^3 [1 - 3.3 (t - 9 \mpi^2) + 5.8
      (t - 9 \mpi^2)^2] & \mbox{\ \ if\ \ } 9 \mpi^2 \leq t < (m_\rho
      + \mpi)^2 \\
      t(1.623 + 10.38/t^2 - 9.32 / t^4 + 0.65/t^6) &
      \mbox{\ \ else\ \ }
   \end{array} \right\}
\ee
(all numbers in appropriate powers of $\unit{GeV}$).

We have studied the influence of a energy-dependent width by comparing
our results to those obtained from
      \be
         \label{eqnf1}
         \Res_{a_1}(t) = \frac{\ma^2 - i \ma \Gamma_{a_1}}%
            {\ma^2 - t - i \ma \Gamma_{a_1}}
      \ee
with a constant width $\ga$.

In order to check our form factor we use it to compute the
radiative decay of the $a_1$. Indeed in the narrow
width approximation, $F_A(\ma^2)$ can be related to
$\gapg$, once the transition
amplitude for $\tau \rightarrow \nu a_1$ is known. Let this
transition be parametrized by the coupling constant $f_{a_1}$
\be
   \mcal(\tau \rightarrow \nu a_1) = G_F \cos \theta_c
   f_{a_1} \epsilon_\mu L^\mu
\ee
Then
\be
   \gapg=
   \frac{\alpha}{48} \frac{\ma^5 \ga^2}{\mpi^2 f_{a_1}^2}
   | F_A(\ma^2) |^2 \left( 1 - \frac{\mpi^2}{\ma^2} \right)^3
\ee
Using equation (\ref{eqnfa}), this can be reexpressed in terms of
$F_A(0)$ as
\be
   \gapg =
   \frac{\alpha}{48} \frac{\ma^7 }{\mpi^2 f_{a_1}^2}
   | F_A(0) |^2 \left( 1 - \frac{\mpi^2}{\ma^2} \right)^3
\ee
{}From the experimental result  with \cite{Rev92}
\be
   \left( \gapg \right)_{exp} = 640 \pm 246 \unit{keV}
\ee
we obtain for the central values
\be
f_{a_1}= 0.09  \unit{GeV}^2
\ee
which has to be compared with
\be
   f_{a_1} = f_\rho = \sqrt{2} \mr \fpi= 0.10 \unit{GeV}^2
\ee
from chiral symmetry and the Kawarabayashi-Suzuki-Riazuddin-%
Fayyazuddin (KSRF)
relation \cite{Kaw66}
or with $f_{a_1}$ extracted from $\tau \rightarrow a_1 \nu $
data:
\be
   f_{a_1}^2 = \frac{\ma^2\mt^2 }{24\pi^2\cos^2\theta_C}
   \frac{1}{\left( 1-\left(\frac{\ma}{\mt}\right)^2\right)^2 %
   \left(1+2\left(\frac{\ma}{\mt}\right)^2\right)}
   \frac{BR(\tau \rightarrow  a_1 \nu )}%
   {BR(\tau \rightarrow e^- \nu \bar{\nu})}
\ee
(where $BR$ denotes branching ratio). With present data,
$BR(\tau \rightarrow
\nu a_1) = 6.8 \%$  \cite{LEP}
 we obtain
 \be
   f_{a_1} = 0.131 \unit{GeV}^2
\ee
from the direct measurement.

So we observe a reasonable agreement between the predicted
and measured radiative decay width of the $a_1$.
%=======================================================================
\subsection{The Vector Form Factor}
The hypothesis of the conserved vector current (CVC) relates
the vector formfactor $F_V$ to the formfactor $\fpg$
of the vertex
$\pi^0 \gamma \gamma^\star$ by
\be
   F_V(t) = - \frac{\mpi \fpg(t)}{\sqrt{2}}
\ee
where $\fpg(t)$ is defined by
\be
   \mcal[\pi^0(p) \longrightarrow \gamma(k) \gamma^\star(q)]
   =  i e^2 \fpg(q^2) \epsilon_{\mu \nu \rho \sigma} k^\mu q^\nu
      \epsilon^\rho(k) \epsilon^\sigma(q)
\ee
The case $q^2 = 0$ corresponds to the decay of a pion into two real photons.

There are three constraints for the form factor $\fpg$:
\begin{itemize}
\item there is a low energy theorem for $\fpg(0)$,
\item the slope of $\fpg(t)$ at $t = 0$ has been measured and
\item there is a perturbative QCD theorem for $q^2 \rightarrow - \infty$.
\end{itemize}

In good agreement with the measured rate for the decay
$\pi^0 \longrightarrow \gamma \gamma$,  $\fpg(0)$ is
predicted by the Adler-Bell-Jackiw anomaly \cite{adler} as
\be
   \fpg(0) = \frac{1}{4 \pi^2 \fpi}
\ee
This yields
\be
   F_V(0) = - 0.0270,
\ee
which is almost within one standard deviation of the experimental value
\cite{Rev92} measured in the radiative pion decay (where $t \approx 0$),
which in our phase convention reads
\be
   F_V(0)_{exp} = - 0.017 \pm 0.008
\ee

Perturbative QCD predicts \cite{Lep79}
\be
    \fpg(q^2 \rightarrow - \infty) \rightarrow - \frac{2 \fpi}{q^2}
\ee

Furthermore the slope of $\fpg$ at $t = 0$ has been measured. Using
the definition
\be
   \fpg(t) = \fpg(0) \left[ 1 + a \frac{t}{m_{\pi^{0}}^2} + O(t^2) \right]
\ee
the experimental value \cite{Rev92} is
\be  \label{eqn2}
   a_{exp} = + 0.0326 \pm 0.0026.
\ee

Assuming $\fpg$ to be dominated by $\rho$ like resonances, we will
try to parameterize $\fpg(t)$ in accordance with the constraints described
above
using the following models:
\begin{itemize}
\item  monopol
      \be
         \fpg(t) = \frac{1}{4 \pi^2 \fpi} \Res_X(t)
      \ee
\item  dipol
      \be
         \fpg(t) = \frac{1}{4 \pi^2 \fpi} \left[\Res_\rho(t) +
            \lambda \Res_\rp(t) \right] \frac{1}{1 + \lambda}
      \ee
\item tripol
      \be
         \fpg(t) = \frac{1}{4 \pi^2 \fpi} \left[\Res_\rho(t) +
            \lambda \Res_\rp(t) + \mu \Res_\rpp(t) \right]
             \frac{1}{1 + \lambda + \mu}
      \ee
\end{itemize}
All these models automatically satisfy the low energy constraint, because
the resonance factors are again normalized according to
\be
   \Res_X(0) = 1
\ee
Furthermore we suppose
\be
    \Res_X(t \rightarrow - \infty) \longrightarrow
    - \frac{m_X^2}{t}
\ee
This is fulfilled by
\be
   \Res_X(t) = \frac{m_X^2}%
      {m_X^2 - t - i m_X \Gamma_{pw}(t)}
\ee
with an energy dependent width corresponding to the
P-wave two body phase space
\be
   \Gamma_{pw}(t) = \left\{ \begin{array}{ll}
       0 & \mbox{\ \ if\ \ } t \leq 4 \mpi^2 \\
       \frac{m_X}{\sqrt{t}} \left( \frac{\sqrt{t - 4 \mpi^2}}%
       {\sqrt{m_X^2 - 4 \mpi^2}} \right)^3 \Gamma_X &
       \mbox{\ \ else\ \ }
       \end{array} \right.
\ee
The form
\be
  \label{eqnf2}
   \Res_X(t) = \frac{m_x^2 - i m_X \Gamma_X}{m_X^2 - t - i m_X
\Gamma_X}
\ee
with a fixed width $\Gamma_X$ will again be considered for comparison.

If we use the physical rho mass $m_\rho = 768 \unit{MeV}$ in the monopol
parametrization,
the high energy limit is 14 \% below the QCD prediction. Brodsky
and Lepage note in \cite{Bro81} that the choice
\be
   m_X = M_{BL} := 2 \sqrt{2} \pi \fpi = 822 \unit{MeV}
\ee
 satisfies the low and the high energy constraint simultaneously
with a
monopole formfactor. The slope parameter $a$ depends on the resonance
mass as follows:
\be
    a = m_{\pi^0}^2 / m_X^2 = \left\{
    \begin{array}{ll}
    0.0309 & \mbox{\ \ \ if\ \ \ } m_X = m_\rho \\
    0.0270 & \mbox{\ \ \ if\ \ \ } m_X = M_{BL}
    \end{array}   \right.
\ee
So $m_X = M_{BL}$ is not a very good approximation in the low energy
region and therefore we will use
\be
   m_X = m_\rho
\ee
for the resonance factor of the monopole parametrization.
This monopole version defines the model $N = 1$.

In the dipol parametrization the parameter $\lambda$ must be
\be
   \lambda = \frac{M_{BL}^2 - m_\rho^2}{\mrp^2 - M_{BL}^2} = 0.0584
\ee
in order to satisfy the QCD theorem. (If not stated otherwise, particle
parameters of \cite{Rev92} are used everywhere.) For the slope
parameter this implies
\be
    a = \frac{m_{\pi_{0}}^2}{1 + \lambda} \left( \frac{1}{\mr^2}
        + \frac{\lambda}{\mrp^2} \right) = 0.0296
\ee
in satisfactory agreement with (\ref{eqn2}).
The dipol formfactor defines the model $N = 2$.

In order to determine the parameters $\lambda$ and $\mu$ of the tripol
model, two more constraints will be considered, viz. the experimental
values for $\grpg$ and $\grppo$. Using the narrow width approximation,
parametrizing
\be
   \mcal(\tau \rightarrow \rho \nu) = G_F \cos \theta_c
   f_\rho \epsilon_\mu L^\mu
\ee
and dividing $F_V$ into
\be
   F_V(t) = F_{V}^{(\rho)} + F_{V}^{(\rho')} + F_{V}^{(\rho'')}
\ee
in an obvious notation, $\grpg$ can be calculated in terms of the
parameters of $F_{V}^{(\rho)}$. The result is
\be
    \grpg = \frac{\alpha}{1536 \pi^4} \frac{\mr^7}{\fpi^2 f_{\rho}^2}
            \left( 1 - \frac{\mpi^2}{\mr^2} \right)^3
            \frac{1}{(1 + \lambda + \mu)^2}.
\ee
We will use $f_\rho$ as given by the KSRF relation,
\be
   f_\rho = \sqrt{2} m_\rho \fpi = 0.10 \unit{GeV}^2.
\ee
Equally well we could extract $f_\rho$ directly from $\tau
\rightarrow \rho \nu$ data:
\be
   f_{\rho}^2 = \frac{\mr^2\mt^2 }{24\pi^2\cos^2\theta_C}
   \frac{1}{\left( 1-\left(\frac{\mr}{\mt}\right)^2\right)^2 %
   \left(1+2\left(\frac{\mr}{\mt}\right)^2\right)}.
   \frac{BR(\tau \rightarrow  \rho \nu )}%
   {BR(\tau \rightarrow e^- \nu \bar{\nu})}
\ee
With present
data \cite {Rev92}  this is
 \be
   f_{\rho} = 0.11 \unit{GeV}^2
\ee
We have checked that the use of this value would not change the final
result significantly.

So for $\lambda = \mu = 0$ (monopole) we obtain
\be
   (\grpg)_{monopole} = 80.5 \unit{keV}
\ee
On the other hand the experimental number is \cite{Rev92}
\be
   (\grpg)_{exp} = (68 \pm 7) \unit{keV}
\ee
and so
\be
    1 + \lambda + \mu = \sqrt{\frac{(\grpg)_{monopole}}{(\grpg)_{exp}}}
    = 1.088
\ee
In order to relate $\grppo$ to the parameters of $F_{V}^{(\rho')}$,
we assume $F_{V}^{(\rho')}$ to be given by the Feynman diagram of
Fig.\ 2. So we need $f_{\rho'}$, $g_{\rho' \omega \pi}$ and
$f_\omega$. $f_{\rho'}$ can be extracted from a fit to $e^+ e^-
\rightarrow 2 \pi$ data \cite{Kue90}, which may proceed through
$\rho$, $\rho'$ or $\rho''$ resonance channels (see Fig.\ 3).
Neglecting the pion mass, $g_{\rho \pi \pi}$ and $g_{\rho' \pi \pi}$
are related by
\be
   \left| \frac{g_{\rho' \pi \pi}}{g_{\rho \pi \pi}} \right|
   = \sqrt{\frac{\Gamma_{\rho' \rightarrow 2 \pi}}
   {\Gamma_{\rho \rightarrow 2 \pi}} \frac{\mr}{\mrp}} = 0.23
\ee
where we used \cite{Rev92}
\be
   \Gamma_{\rho' \rightarrow 2 \pi} = 0.24 \times 0.21 \times 310
   \unit{MeV}
\ee
Let the fit to $e^+ e^- \rightarrow 2 \pi$ be parametrized by
\be
    \left( \Res_{\rho}(t) + \sigma \Res_{\rho'}(t)
    + \rho \Res_{\rho''}(t) \right) \frac{1}{1 + \sigma + \rho}
\ee
Then we have
\be
   \mr^2 : (\sigma \mrp^2) = (f_\rho g_{\rho \pi \pi}) :
   (f_{\rho'} g_{\rho' \pi \pi)}
\ee
Reference \cite{Kue90} gives (see their model $N = 2$, table 1)
\be
   \sigma = - 0.103
\ee
Therefore
\be
   \left| \frac{f_{\rho'}}{f_\rho} \right | =
   \left| \frac{g_{\rho\pi\pi}}{g_{\rho'\pi \pi}} \right|
   \frac{\mrp^2}{\mr^2} \, \sigma = 1.63
\ee
The coupling $g_{\rho' \omega \pi}$ is defined by
\be
   \mcal(\rho'(q) \rightarrow \omega(k) \pi(p)) =
   g_{\rho' \omega \pi} \epsilon_{\mu \nu \rho \sigma}
   \epsilon^\mu(q) \epsilon^\nu(k) k^\rho p^\sigma
\ee
and can be deduced from
\be
   \Gamma_{\rho' \rightarrow \pi \omega} =
   \frac{| g_{\rho' \omega \pi}|^2}{96 \pi} \mrp^3  \left(
   1 - \frac{m_{\omega}^2}{\mrp^2} \right)^3
   = 0.21 \times 310 \unit{MeV}
\ee
as
\be
   | g_{\rho' \omega \pi} | = 4.1 \times 10^{-3} \unit{MeV}^{-1}
\ee
Assuming $f_{\omega} = f_\rho / 3$, comparison of the diagram
of Fig.\ 2 and $F_{V}^{(\rho')}$ then gives
\be
    |\lambda| = \frac{2 \sqrt{2}}{3} (1 + \lambda + \mu)
    \frac{\mpi \fpi^2 \mr^2}{\mrp^2 m_{\omega}^2}
    \frac{| g_{\rho'\omega\pi}|}{F_V(0)}
    \left| \frac{f_{\rho'}}{f_\rho} \right|
    = 0.136
\ee
If $\lambda$ and $1 + \lambda + \mu$ are given with moderate errors,
$\mu$ can only be calculated with a large uncertainty. So instead we
use the QCD limit to express $\mu$ in terms of $\lambda$ and
$1 + \lambda + \mu$ as
\be
   \mu = \frac{1}{\mrpp^2} [ (1 + \lambda + \mu) M_{BL}^2
   - \mr^2 - \lambda \mrp^2]
\ee
We find
\be
\begin{array}{c@{\;\Rightarrow\;}c@{\;\Rightarrow\;}c}
   \lambda = + 0.136  & \mu = - 0.051
   & 1 + \mu + \lambda = 1.085 \\
   \lambda = - 0.136  & \mu = 0.15
   & 1 + \mu + \lambda = 1.014
\end{array}
\ee
So in order be consistent with $1 + \mu + \lambda = 1.088$ from
$\grpg$, $\lambda$ must be $ + 0.136$. The slope parameter $a$ then is
\be
   a = \frac{m_{\pi_0}^2}{1 + \lambda + \mu} \left(
       \frac{1}{\mr^2} + \frac{\lambda}{\mrp^2} + \frac{\mu}{\mrpp^2}
       \right) = 0.0291
\ee
in satisfactory agreement with equation (\ref{eqn2}).

So for the tripol model,
\begin{eqnarray}
   \lambda & = & + 0.136 \nonumber \\
   \mu & = & - 0.051
\end{eqnarray}
is a choice for the two parameters which is consistent with all
the four constraints (the QCD limit, the slope $a$, $\grpg$ and
$\grppo$) and defines the model $N = 3$.

%=======================================================================
\section{Results for the Spectra and the Decay Rate}
We consider as standard the following choices:
\begin{itemize}
\item energy dependent widths in both form factors,
\item model $N = 3$ (tripol) for $F_V$,
\item $F_A(0) = - 0.0116$, and
\item $\Gamma_{a_{1}} = 0.4 \unit{GeV}$.
\end{itemize}
In Fig.\ 4 (a)--(d) the resulting photon spectrum is displayed for
$0.1 \leq x \leq 1$.
For "soft photons" ($x_0 \leq 0.2$) the
internal bremsstrahlung completely dominates (Fig.\ 4(a)). Note that for
very soft photons the multi-photon production rate becomes
important, and so our order $\alpha$ results are not reliable too
close to the infrared divergence $x = 0$. We have checked our internal
bremsstrahlung spectrum against the data produced from PHOTOS
and we obtain excellent agreement, apart from the region of very large
$x$.

The spectrum is
significantly enhanced by structure dependent effects for
hard photons ($x_0 \geq 0.4$)(Fig.\ 4(b)). In Fig.\ 4(c) we show that
the vector resonances dominate the structure dependent part while
in Fig.\ 4(d) we plot the interference between bremsstrahlung and
structure dependent part.

While the integration over the internal bremsstrahlung needs an infrared
cut-off, the structure dependent part does not.
In order to get a feeling on its importance we have performed
the integration over the complete  phase space which yields
\begin{eqnarray}
   \Gamma_{VV} & = & 0.75 (0.86) \cdot 10^{-3} \gnr \nonumber \\
   \Gamma_{AA} & = & 0.33 \cdot 10^{-3} \gnr \nonumber \\
   \Gamma_{VA} & = & 0 \nonumber \\
   \Rightarrow \Gamma_{SD} & = & 1.08 (1.20)  \cdot 10^{-3} \gnr
\end{eqnarray}
using the tripol  parametrization for $F_V$. The numbers in
parentheses are the predictions of the monopole fit.

With the value of \cite{Rev92} for the branching ratio
$\br(\tau \rightarrow \pi\nu)$ this translates into
\begin{eqnarray}
   \br_{VV}(\tau \rightarrow \nu \pi \gamma) & = &
   0.87 (1.00) \cdot 10^{-4} \nonumber \\
   \br_{AA}(\tau \rightarrow \nu \pi \gamma) & = &
   0.38 \cdot 10^{-4}
\end{eqnarray}
These values may be compared to narrow width estimates:
taking into account the lowest lying resonances
$\rho$, $a_1$ we obtain
$$
   \br(\tau \rightarrow \nu \rho \rightarrow \nu \pi \gamma)
   \approx \br(\tau \rightarrow \nu \rho) \times
     \br(\rho \rightarrow \pi \gamma) \approx
     22 \% \times 4.5 \cdot 10^{-4} = 0.99 \cdot 10^{-4}
$$ \be
    \br(\tau \rightarrow \nu a_1 \rightarrow \nu \pi \gamma)
    \approx \br(\tau \rightarrow \nu a_1) \times
    \br(a_1 \rightarrow \pi \gamma) \approx
    6.8 \% \times 1.6 \cdot 10^{-3} = 1.01 \cdot 10^{-4}
\ee
Here we have used the central value
\be
   \br(a_1 \rightarrow \pi \gamma) = \frac{640 \unit{keV}}%
      {400 \unit{MeV}} = 1.6 \cdot 10^{-3}
\ee
So for both the vector and the axial vector channel the narrow width
approximation and the
picture of a decay chain $\tau \rightarrow \nu \rho\, (a_1)$, $\rho\,
(a_1)\, \rightarrow \pi \gamma$ give reasonable results (do not
forget the large error on $BR(a_1\to \pi\gamma)$).

In tables \ref{tab0a} and \ref{tab0b} we display
for two values of the photon energy cut
how the different parts contribute to the total rate (using
the standard parameters).
Note that the only term involving $F_V$ which is of importance
is $\Gamma_{VV}$, while for $F_A$ the internal bremsstrahlung-%
structure dependent interference part $\Gamma_{IB-A}$ is of the same size
as $\Gamma_{AA}$. While the vector-axial interference $\Gamma_{VA}$
can always be neglected, the internal bremsstrahlung-structure
dependent interference $\Gamma_{INT}$ is rather small but not
negligible.

\begin{table}
%\centerline{\bf Table 1}
\caption{Contribution of the different parts to the total rate,
using $E_\gamma \geq 400 \unit{MeV}$}
\begin{eqnarray*}
   \hline \hline
   \Gamma_{IB} (x_0)& = & 1.48 \cdot 10^{-3} \gnr \nonumber \\
   \left. \begin{array}{lll}
       \Gamma_{VV}(x_0) & = & 0.55 \cdot 10^{-3} \gnr  \\
       \Gamma_{VA}(x_0) & = & 0.01 \cdot 10^{-3} \gnr  \\
       \Gamma_{AA}(x_0) & = & 0.31 \cdot 10^{-3} \gnr
   \end{array} \right\}
       \Gamma_{SD}(x_0) & = & 0.87 \cdot 10^{-3} \gnr \nonumber \\
   \left. \begin{array}{lll}
       \Gamma_{IB-V}(x_0) & = & 0.07  \cdot 10^{-3} \gnr  \\
       \Gamma_{IB-A}(x_0) & = & 0.35  \cdot 10^{-3} \gnr  \\
   \end{array} \right\}
       \Gamma_{INT}(x_0) & = & 0.42  \cdot 10^{-3} \gnr \nonumber \\
      \mbox{\ \ } \nonumber \\
   \Rightarrow \Gamma_{total}(x_0) & = & 2.76  \cdot 10^{-3} \gnr.\\
   \hline \hline
\end{eqnarray*}
\label{tab0a}
\end{table}
\begin{table}
%
%\centerline{ \bf Table 2}
\caption{Contribution of the different parts to the total rate,
using $E_\gamma \geq 50 \unit{MeV}$}
\begin{eqnarray*}
   \hline \hline
   \Gamma_{IB}(x_0) & = & 13.1 \cdot 10^{-3} \gnr \nonumber \\
   \left. \begin{array}{lll}
       \Gamma_{VV}(x_0) & = & 0.75 \cdot 10^{-3} \gnr  \\
       \Gamma_{VA}(x_0) & = & 0.00 \cdot 10^{-3} \gnr  \\
       \Gamma_{AA}(x_0) & = & 0.33 \cdot 10^{-3} \gnr
   \end{array} \right\}
       \Gamma_{SD}(x_0) & = & 1.08 \cdot 10^{-3} \gnr \nonumber \\
   \left. \begin{array}{lll}
       \Gamma_{IB-V}(x_0) & = & 0.05  \cdot 10^{-3} \gnr  \\
       \Gamma_{IB-A}(x_0) & = & 0.57  \cdot 10^{-3} \gnr  \\
   \end{array} \right\}
       \Gamma_{INT}(x_0) & = & 0.62  \cdot 10^{-3} \gnr \nonumber \\
      \mbox{\ \ } \nonumber \\
   \Rightarrow \Gamma_{total}(x_0) & = & 14.8 \cdot 10^{-3} \gnr.\\
   \hline \hline
\end{eqnarray*}
\label{tab0b}
\end{table}

In table \ref{tab1} the dependence of the total rate on the variation
of the parameters is displayed for $E_\gamma \geq 400 \unit{MeV}$.
It turns out that the uncertainties in the vector form factor $F_V$
(whether we use a monopole, dipole or tripol form and whether we
assume fixed or energy dependent width) only have rather little impact
on the total rate. Its uncertainty is dominated by the ambiguities of
the axial form factor $F_A$, in which connection the value for the $a_1$
width
and the normalization error at $t = 0$ are of comparable importance.
\begin{table}
%\centerline{\bf Table 3}
   \caption{Dependence of the total rate on variation of the
      parameters. Standard choices are implied wherever not
      stated otherwise. $ E_\gamma \geq 400\unit{MeV}$. }
$$
   \begin{array}{cl}
      \mbox{Variation} & \Gamma_{total} \left( 10^{-3} \gnr
      \right) \\
      \hline \hline
      N = 1 \mbox{\ (monopol)} & 2.88 \\
      N = 2 \mbox{\ (dipol)}   & 2.77 \\
      N = 3 \mbox{\ (tripol)}  & 2.76 \mbox{\ (standard)} \\
      \hline
      \mbox{fixed widths for the $\rho$'s} & 2.80 \\
      \hline
      \ga = 0.6 \unit{GeV} \mbox{, variable width\ \ } & 2.60\\
      \ga = 0.4 \unit{GeV} \mbox{, variable width\ \ } & 2.76
      \mbox{\ (standard)} \\
      \ga = 0.25 \unit{GeV} \mbox{, variable width\ \ } & 3.05 \\
      \hline
      \ga = 0.6 \unit{GeV} \mbox{, fixed width\ \ } & 2.64 \\
      \ga = 0.4 \unit{GeV} \mbox{, fixed width\ \ } & 2.79 \\
      \ga = 0.25 \unit{GeV} \mbox{, fixed width\ \ } & 3.07 \\
      \hline
      F_A(0) = - 0.0100 & 2.63 \\
      F_A(0) = - 0.0116 & 2.76 \mbox{\ (standard)} \\
      F_A(0) = - 0.0132 & 2.90 \\
      \hline
      F_A(0) = + 0.0422 & 4.94 \\
      \hline
      F_A(t) \equiv F_A(0) ,\, F_V(t) \equiv F_V(0) &
      1.25\\
      \hline \hline
   \end{array}
$$
   \label{tab1}
\end{table}

Note that choice of the positive sign solution for $F_A$ almost
doubles the integrated rate, and so by measuring $\tau \to\nu\pi
\gamma$ a clear decision for one solution is possible.

In the last column of this table we display the result of an
integration with the formfactors fixed at their values at $t = 0$
without Breit-Wigner resonances. It turns out that the Breit-Wigner
enhancement substantially increases the total rate.

Fig.\ 5 displays the integrated decay rate in variation with the photon
energy cut $E_0$ ($E_\gamma \geq E_0$).
We give the prediction
using the standard parameter set, a lower and upper limit
and the contribution from pure internal
bremsstrahlung.
The lower limit is obtained using
\begin{itemize}
\item $N = 3$, tripole form of $F_V(t)$,
\item $\ga =  0.6 \unit{GeV}$,
\item variable widths for the formfactors, and
\item $F_A(0) = - 0.0100$
\end{itemize}
the upper limit corresponds to
the following choice:
\begin{itemize}
\item $N = 1$, monopole form of $F_V(t)$,
\item $\ga =  0.25 \unit{GeV}$,
\item fixed widths for the formfactors, and
\item $F_A(0) = - 0.0132$.
\end{itemize}
%****************
%*****************

In Figs.\ 6 (a) --- (d) we show the pion-photon invariant mass spectrum.
We find that a much better seperation between
internal bremsstrahlung and structure dependent effects is obtained here
(Fig. 6 (a) and (b)),
as compared with the photon spectrum.
So this spectrum (in the region $\sqrt{t} \geq 700 \unit{MeV}$)
is an observable that is much better suited for studying the form factors
than the photon spectrum is.
It turns out that with the $a_1$ width of $\ga = 400\unit{MeV}$,
the $\rho$ is the only resonance visible in the invariant mass spectrum.
The total structure dependent (SD) spectrum is the sum of the vector and
axial vector Breit-Wigners (VV and AA),
as the vector-axial interference (VA) vanishes
in the invariant mass spectrum after integration over the other kinematical
variable. So in the SD spectrum there is a very soft and broad $a_1$ resonance
bump (Fig.\ 6 (c)). The internal bremsstrahlung-structure dependent inter%
ference radiation (INT) near the $a_1$
is dominated by the axial part (IB-A). It changes sign
at the $a_1$ mass and rises strongly with decreasing $\sqrt{t}$ below
$\ma$ (Fig.\ 6 (d)). The internal bremsstrahlung (IB) also rises below
$\ma$, and so the $a_1$ bump is completely erased in the total spectrum
(Fig. 6 (b)).

In Fig.\ 7 we plot the spectrum in the $a_1$ mass region using $\ga =
250$, $400$ and $600 \unit{MeV}$. The result is that
the $a_1$ resonance peak is visible if and only if
the $a_1$ width is small ($\ga \approx 250 \unit{MeV}$).

We close this section by discussing the integrated rate with an
invariant mass cut.
For this invariant mass cut $\sqrt{t_0}$ we suggest three different
values: The rather low value of $400 \unit{MeV}$, which gives a rather high
integrated rate, the value $700 \unit{MeV}$, which suppresses the internal
bremsstrahlung without loosing much structure dependent radiation and
$1000 \unit{MeV}$ in order to focus on the axial channel (see
Tab. \ref{tabt1}).

For the total rate we find
\begin{eqnarray}
   \Gamma(\sqrt{t_0} = 400 \unit{MeV}) & = & 4.83\, (4.55 \dots 5.62)
   10^{-3} \gnr
   \nonumber \\
   \Gamma(\sqrt{t_0} = 700 \unit{MeV}) & = & 2.17\, (1.91 \dots 2.95)
   10^{-3} \gnr
   \nonumber \\
   \Gamma(\sqrt{t_0} = 1000 \unit{MeV}) & = & 0.72\, (0.52 \dots 1.28)
   10^{-3} \gnr
\end{eqnarray}
where the central values correspond to the standard parameter set
and the lower and upper limit are obtained with the same
parameter variations as above.

If we vary $\ga$ from $250 \unit{MeV}$ to $600 \unit{MeV}$ and keep
the other parameters fixed at their standard values, the total rate
$\Gamma(\sqrt{t_0} = 1000 \unit{MeV})$
varies from $1.00$ to $0.56$ $10^{-3} \gnr$, if we vary $F_A(0)$
from $-0.0100$ to $-0.0132$, the total rate varies from $0.64$
to $0.81$ $10^{-3} \gnr$. So because of the experimental uncertainty
of $F_A(0)$ it is difficult to get much information on the $a_1$ width
by measuring the integrated rate and the invariant mass spectrum
should be used, as discussed above.

\begin{table}
\caption{Contributions to the integrated rate $\Gamma(\protect\sqrt{t_0})$
using the standard parameter set
(all numbers for the rate in units of $10^{-3} \gnr$)
}
$$ \begin{array}{ccccccc}
   \sqrt{t_0} (\unit{MeV}) & \Gamma_{IB} & \Gamma_{VV} & \Gamma_{AA}
   & \Gamma_{IB-V} & \Gamma_{IB-A} & \Gamma_{total} \\
   \hline \hline
    400 & 3.19 & 0.74 & 0.33 & 0.05 & 0.51 & 4.83 \\
    700 & 0.77 & 0.61 & 0.33 & 0.13 & 0.33 & 2.17 \\
   1000 & 0.19 & 0.09 & 0.28 & 0.07 & 0.08 & 0.72 \\
   \hline \hline
\end{array} $$
\label{tabt1}
\end{table}

%=======================================================================
\section{Estimation of the Decay $\tau \to \nu K \gamma$}
Compared with the pionic mode, the decay $\tau \to \nu K \gamma$
is both Cabbibo and phase space suppressed, i.e. it is even more
rare. So a reasonable estimate for the total rate $\Gamma_K(x_0)$
will be sufficient.

For $\tau \to \nu K \gamma$, in Secs.\ 2 and 3 the following
substitutions must be made:
\begin{eqnarray*}
   \cos \theta_c & \longrightarrow & \sin \theta_c \\
   \fpi & \longrightarrow & f_K = 113 \unit{MeV}   \\
   F_{V/A}(t) & \longrightarrow & F_{V/A}^{(K)}(t) \\
   \mpi & \longrightarrow & m_K \\
   \gnr & \longrightarrow & \Gamma_{\tau \rightarrow \nu K}.
\end{eqnarray*}
Flavor symmetry implies the following relations for the
formfactors at $t = 0$:
\begin{eqnarray}
   F_{A}^{(K)} (0) & = & \frac{m_K}{\mpi} F_A(0) = - 0.0410,
   \nonumber \\
   F_{V}^{(K)} (0) & = & \frac{m_K}{\mpi} F_V(0) = - 0.0955.
\end{eqnarray}
As far as the quantum numbers are concerned, the resonances
$K^\star(892)$, $K^\star(1410)$ and $K^\star(1680)$ could
contribute in the vector channel and $K_1(1270)$ and $K_1(1400)$
in the axial one. The rates $\Gamma(K_1(1400) \to K \rho, K \omega)$
and $\Gamma(K^\star(1410) \to K \rho)$, however, are compatible with
zero \cite{Rev92}, and so, assuming vector meson dominance, their
contribution to $\tau \to \nu K \gamma$ may be neglected.
In the case of the pionic mode, the contribution of the $\rho''$
was found to be small. Therefore we assume that in the kaonic
mode the contribution of the $K^\star(1680)$
is small also and approximate it by zero.
So we are left with the $K^\star(892)$ and the $K_1(1270)$ and
a monopole form for both form factors:
\begin{eqnarray}
   F_{A}^{(K)} (t) & = & F_{A}^{(K)}(0) \Res_{K_1(1270)} (t)
   \nonumber \\
   F_{V}^{(K)} (t) & = & F_{V}^{(K)}(0) \Res_{K^\star(892)} (t).
\end{eqnarray}
For simplicity we use constant widths as in equations (\ref{eqnf1}) and
(\ref{eqnf2}) for both the $K^\star$ and the $K_1$.

The resulting integrated decay rate $\Gamma_K(E_0)$ is plotted
in Fig.\ 8 in dependence on the photon energy cut $E_0$. Note
that here the structure dependent effects are much  more important
than in the pionic case. This is understood easily, because the
higher mass of the kaon suppresses the internal bremsstrahlung.
So the measurement of structure dependent effects is fairly easy
is this decay mode.

We have also performed the integration using a kaon-photon invariant
mass cut. The results are:
\begin{eqnarray}
   \Gamma_K(\sqrt{t_0} = 800 \unit{MeV}) & = & 3.58 \cdot 10^{-3}
   \Gamma_{\tau \rightarrow \nu K}
  \nonumber \\
   \Gamma_K(\sqrt{t_0} = 1200 \unit{MeV}) & = & 0.91 \cdot 10^{-3}
   \Gamma_{\tau \rightarrow \nu K}
\end{eqnarray}
%=======================================================================
\section{Summary and Discussion}
We have calculated the radiative one pseudoscalar decay modes
of the tau lepton. Because of the infrared divergence of the
internal bremsstrahlung the integrated decay rates depend on
a photon energy cut $E_0$ or equivalently on a meson-photon
invariant mass cut $t_0$.
The branching ratio for the pionic
mode is found to be of the order of $10^{-3}$ $(10^{-4})$,
using a rather low (high) value for $E_0$. For the kaonic mode,
the branching ratio is of the order of $10^{-5}$. So at least
the pionic radiative mode should be observable presently.

The rate for $\tau \to \nu \pi \gamma$ is dominated by the internal
bremsstrahlung, and so, in order to see structure dependent effects,
sufficient statistics and/or investigation of the invariant mass
spectrum rather than the photon energy spectrum
are needed. Because of the higher
mass of the kaon, the observation of ``structure'' will be easier
in the decay $\tau \to \nu K \gamma$.

Our basic assumption in the parametrization of the form factors
$F_V$ and $F_A$ was that they are dominated by hadron resonances
with suitable quantum numbers.

In order to parametrize $F_V$ in the pionic case,
we then used the CVC-hypothesis
to relate $F_V(t)$ to $F_{\pi \gamma}(t)$ and the QCD-theorem
on $\lim_{t \to \infty} \left( t F_{\pi\gamma(t)} \right)$
and assumed a saturation
of the QCD-theorem by the $\rho$-like resonances of lowest mass.
While the other assumptions seem rather safe, there is no obvious
{\em a priori} justification for the last one. We find, however,
that the $\rho(770)$ alone saturates the QCD-theorem at the $86 \%$
level and that the inclusion of the $\rho'$ and the $\rho''$ changes
the picture only little, which supports our assumption {\em a
posteriori}. Then if this saturation is believed in, the remaining
ambiguities in the parametrization of $F_V(t)$ have only little
impact on the decay rate for $\tau\to\nu\pi\gamma$.

In the axial channel the $a_1$ is the only known resonance which
has the correct quantum numbers. Still $F_A(t)$ dominates the
uncertainty of the prediction for $\tau\to\nu\pi\gamma$,
because neither $F_A(0)$ from radiative pion decay nor
the width of the $a_1$ are known very well experimentally.
In this paper we took the attitude that the negative sign solution
for $F_A$ (corresponding to $\gamma \approx 1/2$) is the physical one.
Measurements from pions at rest
also allow for a solution with $\gamma \approx -2.5$.
We showed that the prediction for $\tau\to\nu\pi\gamma$ changes
very significantly if $\gamma \approx -2.5$, and so a measurement of
this decay will allow for a clear independent check
of $\gamma \approx 1/2$.
Furthermore we showed that by measuring the pion-photon invariant
mass spectrum in the $a_1$ region a decision for a small or large
$a_1$ width can be made.
So it will be very interesting to see
whether the measurement of $\tau\to\nu\pi\gamma$ will support
the large $\ga$ (of $400 \dots 600 \unit{MeV}$)
as measured in the channel $\tau\to\nu 3 \pi$ or a small $a_1$ width
(of about $300 \unit{MeV}$),
as prefered by hadronically produced $a_1$.
% With this solution we obtain
%for instance $\Gamma(400 MeV)=\ 4.9 \times 10^{-3}\times\gnr$
%as compared to $2.76 \times 10^{-3}\times\gnr$.

So we believe that
once a high statistics sample of $\tau\to\nu\pi\gamma$
is available we will learn much about low energy physics.

In a forthcoming paper we will address the question of soft photons
and present the total decay rate $\Gamma(\tau\to \nu_\tau+\pi+(\gamma)
)$.
%=======================================================================
\section*{Acknowledgement}
M.~F. would like to thank J. Bijnens for an illuminating
discussion and M. Schmidtler for producing the spectrum using
PHOTOS.

%=======================================================================

\end{document}